\documentclass[12pt]{iopart}

\usepackage[dvips]{graphicx}
\usepackage{lineno}

\begin{document}

\title{
  STAR Highlights
}

\author{Hiroshi Masui for the STAR collaboration}

\address{
  Lawrence Berkeley National Laboratory,
    Nuclear Science Division,
    1 Cyclotron Road,
    Berkeley, CA 94720
}
\ead{HMasui@lbl.gov}
\begin{abstract}
  We report selected results from STAR collaboration at RHIC,
  focusing on jet-hadron and jet-like correlations,
  quarkonium suppression and collectivity,
  di-electron spectrum in both p~+~p and Au~+~Au,
  and higher moments of net-protons as well as azimuthal anisotropy
  from RHIC Beam Energy Scan program.
\end{abstract}


\section{Introduction}

  The main goals at the STAR experiment for ultra-relativistic heavy ion collisions 
  is to study the structure of QCD phase diagram. In order to achieve this question,
  the STAR experiment has two main programs in heavy ion collisions;
  study the medium properties and eventually identify the equation of state of partonic matter,
  and search for the QCD critical point through RHIC Beam Energy Scan (BES).

  This article is organized as follows. In Section~\ref{sec:jets}, recent results on 
  jet-hadron and jet-like correlations are presented, which can be used to 
  understand underlying mechanisms of jet-quenching and hadron productions at higher $p_T$.
  The suppression and collectivity for heavy flavors are discussed through the measurement
  of the transverse momentum $p_T$ spectra, nuclear modification factor $R_{AA}$
  and azimuthal anisotropy $v_2$ in Section~\ref{sec:heavy}.
  In Section~\ref{sec:em}, the first STAR measurement of di-electron spectrum in 
  both p~+~p and Au~+~Au are presented.
  Section~\ref{sec:bes} presents selected results from BES about
  higher moments of net-protons and second harmonic of azimuthal anisotropy $v_2$
  for identified hadrons

\section{Jet-hadron and jet-like correlations}
\label{sec:jets}

  \begin{figure}[htbp]
  \begin{tabular}{cc}
    \begin{minipage}[htbp]{0.50\textwidth}
      \includegraphics[width=\linewidth]{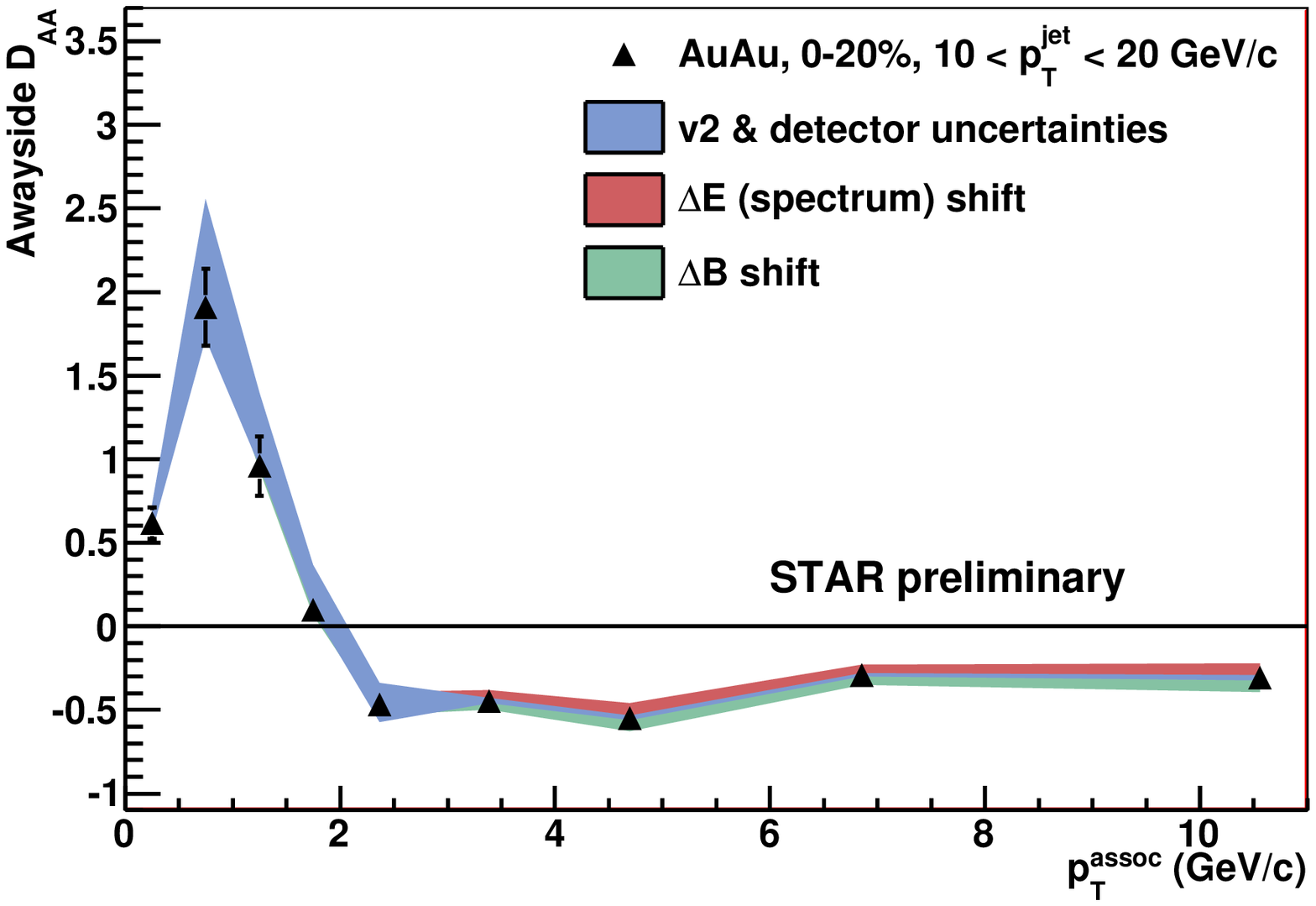}
    \end{minipage}
    \hfill
    \begin{minipage}[htbp]{0.50\textwidth}
      \includegraphics[width=\linewidth]{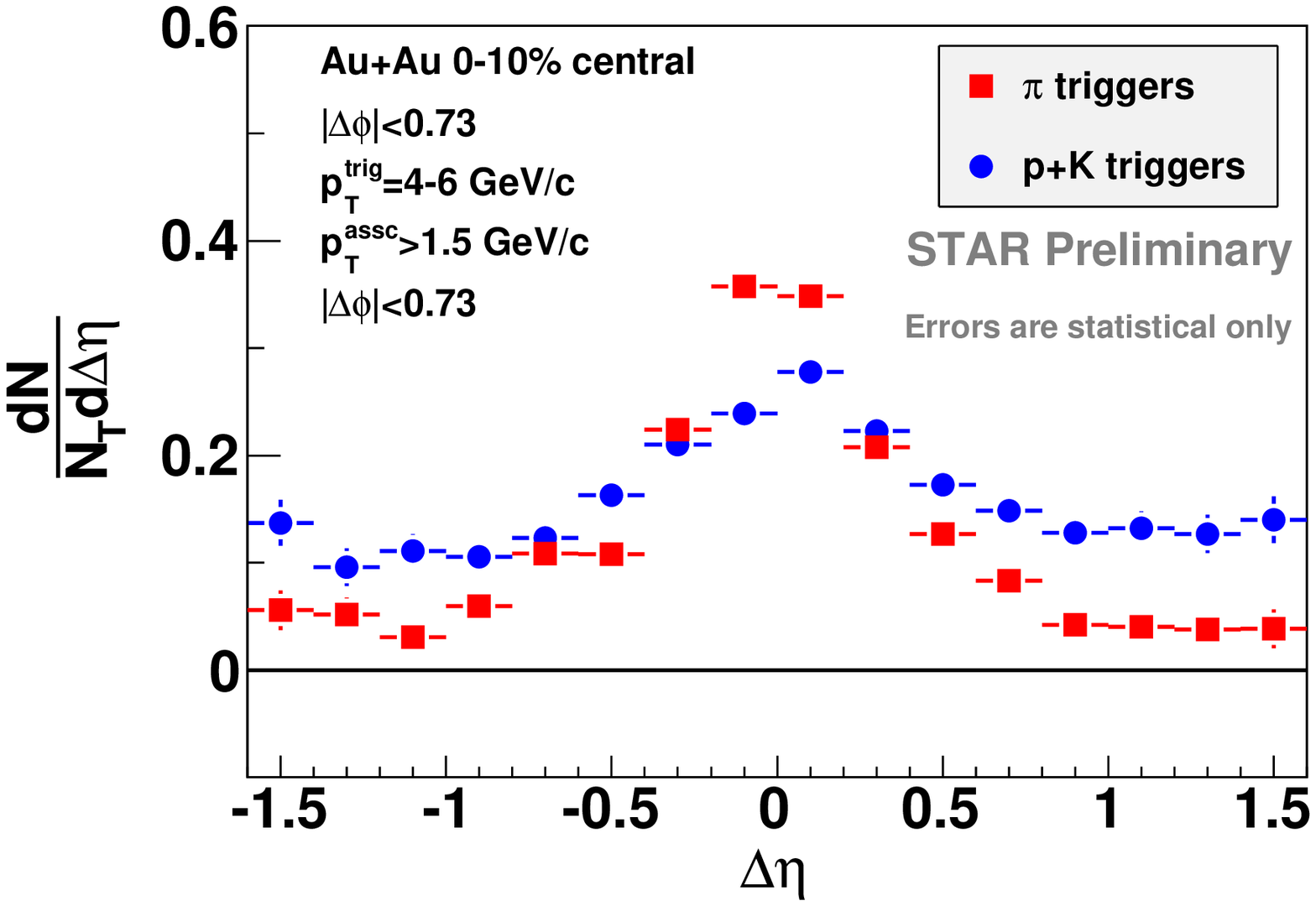}
    \end{minipage}
  \end{tabular}
      \caption{ (Color online)
        (Right) Away side energy balance $D_{AA}$ as a function of associated
        hadron $p_T$ ($p_T^{assoc}$)
        in 0-20\% Au~+~Au collisions at $\sqrt{s_{NN}}$~=~200~GeV,
        where the bands represent systematic uncertainties arising from several 
        different sources quoted in the figure.
        (Left) $\Delta\eta$ distributions for $\pi$ (squares) and $p+K$ (circles) triggered 
        hadron correlation in central 0-10\% Au~+~Au collisions at $\sqrt{s_{NN}}$~=~200~GeV.
        Only statistical errors are shown.
      }
      \label{fig:fig1}
  \end{figure}

    Measurements of $R_{AA}$ as well as di-hadron azimuthal correlations 
  for high $p_T$ inclusive charged hadrons have indicated large partonic energy loss in the medium
  at RHIC~\cite{Adams:2003kv, Adler:2002tq}.  Jet-hadron correlation with full jet reconstructions can 
  provide direct access to the energy of partons in order to understand the underlying mechanisms how partons lose energy in the medium.
  Figure~\ref{fig:fig1} (left) shows the away side ($\Delta\phi \sim \pi$) energy balance $D_{AA}$ as a function 
  of associated hadron $p_T^{assoc}$ in 0-20\% Au~+~Au collisions at $\sqrt{s_{NN}}$~=~200~GeV~\cite{Alice:QM2011}.
  The $D_{AA}$ is defined by 
  \begin{equation}
    D_{AA}(p_T^{assoc}) = Y_{AA}(p_T^{assoc}) \cdot p_{T,AA}^{assoc} - Y_{pp}(p_T^{assoc}) \cdot p_{T,pp}^{assoc},
  \end{equation}
  where $Y(p_T^{assoc})$ denotes the away side jet yields.
  In central 0-20\%, the away side jets are significantly enhanced at low $p_T^{assoc}$, and suppressed at high $p_T^{assoc}$. 
  The integrated energy balance for the away side jets $\Delta B = 1.6^{+1.6+0.5}_{-0.4-0.4}$~GeV/$c$
  reveals a large part of high $p_T^{assoc}$ suppression is balanced by low $p_T^{assoc}$ enhancement.
  This result together with the observation of significant broadening of awayside jets in Au~+~Au compared to p~+~p
  seems to be consistent with the radiative energy loss picture.

  The $p/\pi$ ratio at intermediate $p_T$ ($2 < p_T < 6$~GeV/$c$) have been qualitatively described 
  by recombination models in the most central Au~+~Au collisions~\cite{Adler:2002tq},
  which indicates that coalescence or recombination process would play an important role for hadron productions 
  at intermediate $p_T$. More differential study by utilizing the di-hadron correlation with identified 
  particles would provide further insights into the hadron production mechanisms at intermediate $p_T$.
  Figure~\ref{fig:fig1} (right) shows the $\Delta\eta$ distributions for identified $\pi$ and $p+K$ 
  in $4 < p_T^{trig} < 6$~GeV/$c$ associated with inclusive charged hadrons in $1.5 < p_T^{assoc} < p_T^{trig}$~GeV/$c$ 
  in the most central 0-10\% Au~+~Au collisions~\cite{Kolja:QM2011}, where $\eta$ denotes the pseudorapidity. 
  The $\pi$ and $p+K$ are identified by their specific energy loss in the Time Projectioin Chamber (TPC).
  The shoulder structures in the large $\Delta\eta$ (ridge) is 
  higher for $p+K$ trigger than that for $\pi$ trigger. The near side peak is higher for $\pi$ trigger 
  compared to that for $p+K$ trigger, which is similar in both d~+~Au and Au~+~Au collisions.
  The similarity of near side peak in d + Au and Au~+~Au collisions pose a challenge to the naive 
  recombination picture, where the dilution of near side peak with thermalized partons 
  would be expected if the recombination is the dominant mechanism of hadron
  productions at intermediate $p_T$.

\section{Heavy flavors}
\label{sec:heavy}

  The suppression of $J/\psi$ due to color screening of $c\bar{c}$ pairs
  is expected to be one of the unambiguous signature of QGP formation~\cite{Matsui:1986dk}.
  Measurements of nuclear modification factor $R_{AA}$ for $J/\psi$
  in $\sqrt{s_{NN}}$~=~200~GeV Au~+~Au collisions at RHIC
  show the similar suppression pattern as at SPS as a function of centrality~\cite{Adare:2006ns},
  while the $R_{AA}$ for high $p_T$ ($p_T > $~5~GeV/$c$) $J/\psi$ is consistent with 
  no suppression in Cu + Cu collisions~\cite{Abelev:2009qaa}.
  It has been reported that the suppression of $\Upsilon$ at RHIC could provide
  large sensitivity to color screening in the QGP because contributions from
  regeneration~\cite{Grandchamp:2005yw} and absorption by hadronic co-movers 
  are negligible~\cite{Lin:2000ke}. The quarkonium suppression
  would be one of the probes to study the properties of the QGP as well as to 
  understand their production mechanisms.

  \begin{figure}[htbp]
  \begin{tabular}{cc}
    \begin{minipage}[htbp]{0.50\textwidth}
      \includegraphics[width=\linewidth]{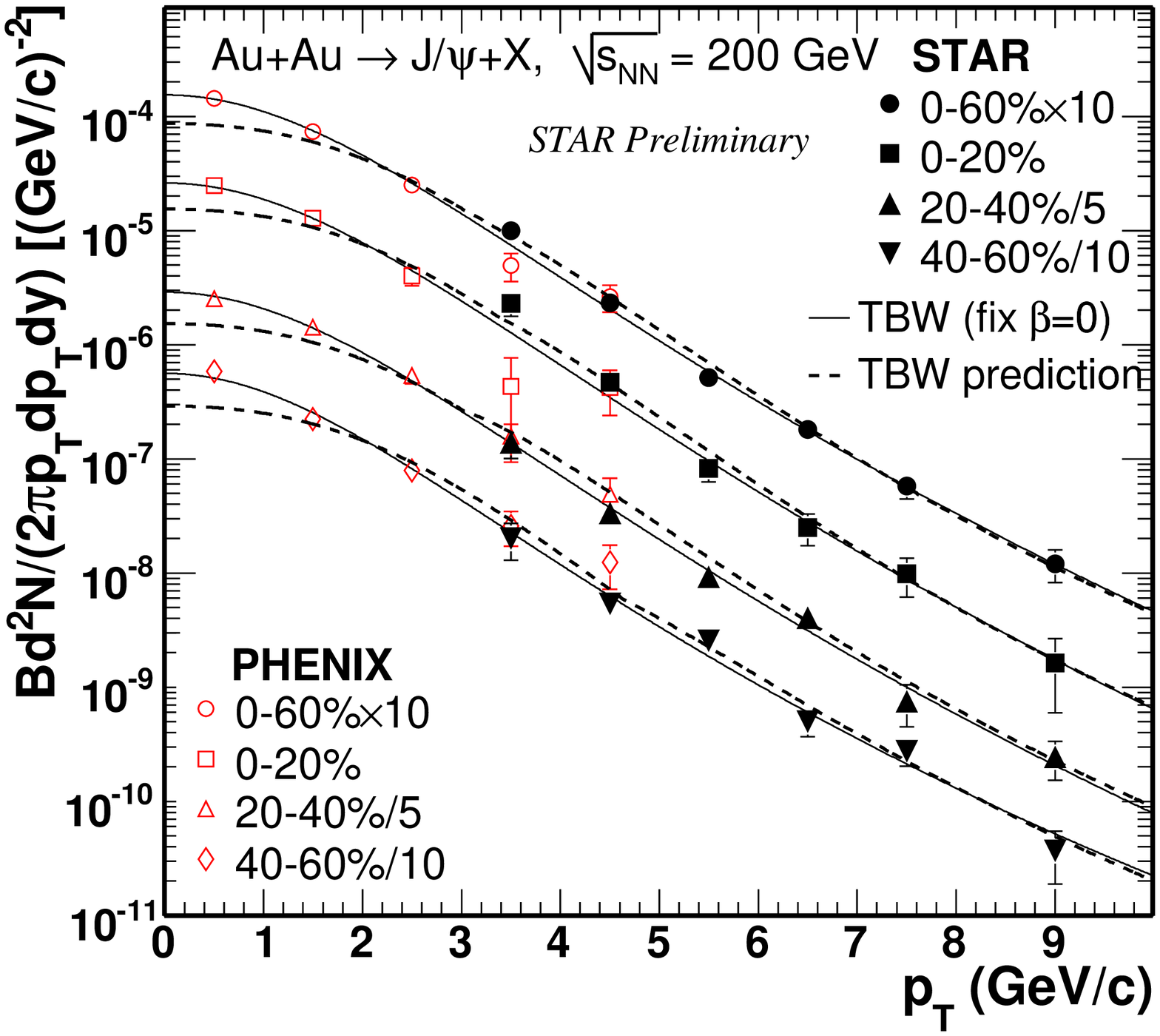}
    \end{minipage}
    \hfill
    \begin{minipage}[htbp]{0.50\textwidth}
      \includegraphics[width=\linewidth]{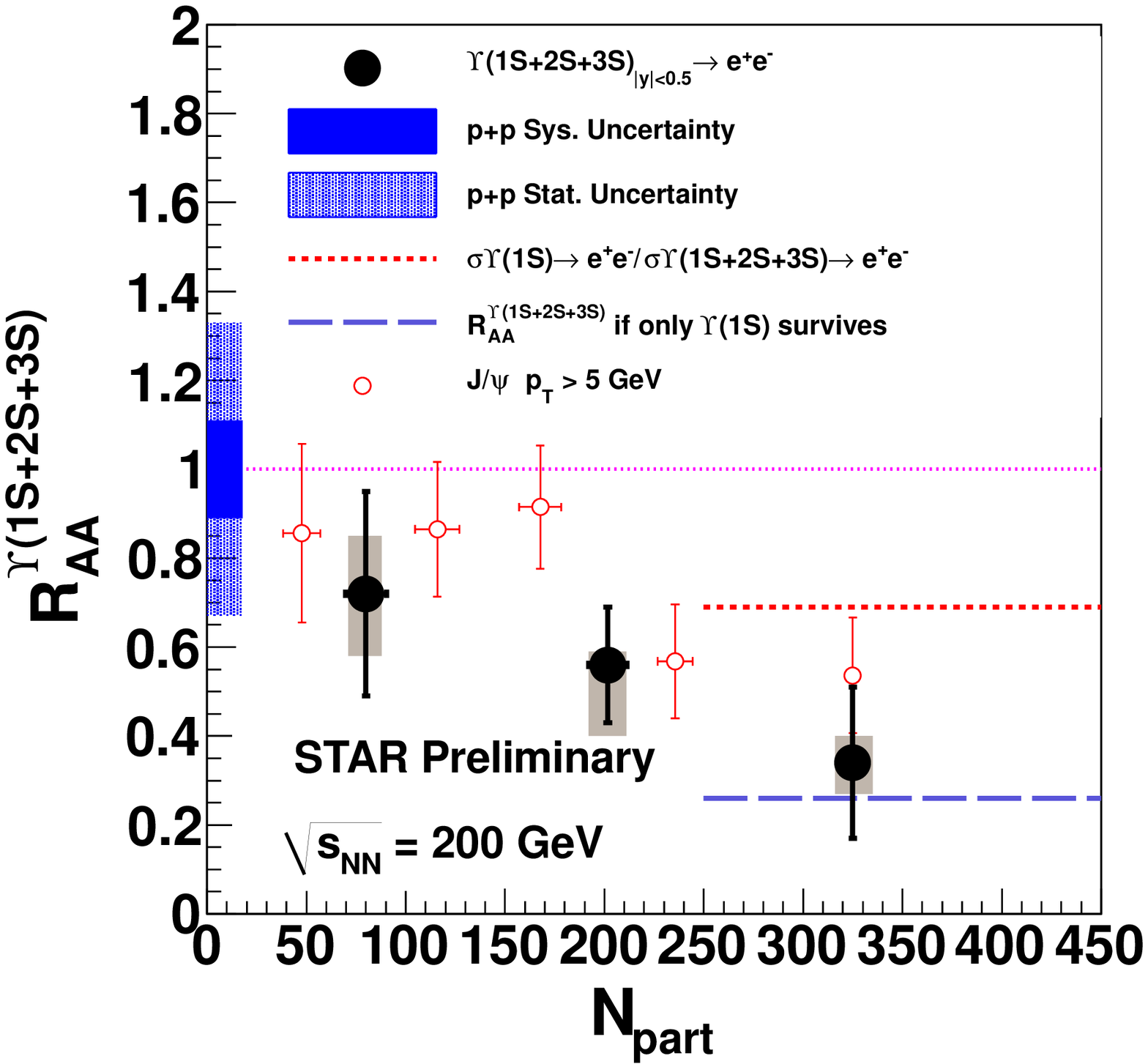}
    \end{minipage}
  \end{tabular}
      \caption{ (Color online)
        (Right) Invariant yields of $J/\psi$ as a function of $p_T$ for several different centrality bins 
        shown by solid symbols in Au~+~Au collisions at $\sqrt{s_{NN}}$~=~200~GeV. 
        Open symbols are published PHENIX results for comparison~\cite{Adare:2006ns}.
        Lines are calculations from Tsallis-Blast-Wave model~\cite{Tang:2011xq}.
        (Left) $\Upsilon$(1S+2S+3S) $R_{AA}$ as a function of $N_{part}$ compared to the 
        $J/\psi$ $R_{AA}$ for $p_T >$~5~GeV/$c$ in Au~+~Au collisions at $\sqrt{s_{NN}}$~=~200~GeV.
        The shaded bands at $N_{part}$~=~0 represent statistical and systematic uncertainties
        in p~+~p measurements.
      }
      \label{fig:fig2}
  \end{figure}

  Figure~\ref{fig:fig2} (left) shows the invariant yields of $J/\psi$ as a function of $p_T$
  in Au~+~Au collisions at $\sqrt{s_{NN}}$~=~200~GeV for several different centrality bins.
  High statistics data in year 2010 allows us to extend the $p_T$ reach of $J/\psi$ spectra
  out to 10~GeV/$c$~\cite{Tang:QM2011}.  Figure~\ref{fig:fig2} (right) shows the $R_{AA}$ for high $p_T$ $J/\psi$
  ($p_T > $~5~GeV/$c$) and $\Upsilon$(1S+2S+3S) as a function of 
  number of participants $N_{part}$ in Au~+~Au collisions at $\sqrt{s_{NN}}$~=~200~GeV.
  The high $p_T$ $J/\psi$ shows suppression in central collisions, while
  the magnitude of $R_{AA}$ is higher in higher $p_T$ compared to the measurements 
  at low $p_T$ from~\cite{Adare:2006ns}.
  The $\Upsilon$(1S+2S+3S) $R_{AA}$ shows similar suppression pattern as high $p_T$ $J/\psi$.
  The $R_{AA}$ in the most central 0-10\% indicates the suppression of $\Upsilon$,
  which is more than 3~$\sigma$ away from unity, with systematic uncertainties in p~+~p~\cite{Reed:QM2011}.

  \begin{figure}[htbp]
    \begin{center}
      \includegraphics[width=0.70\linewidth]{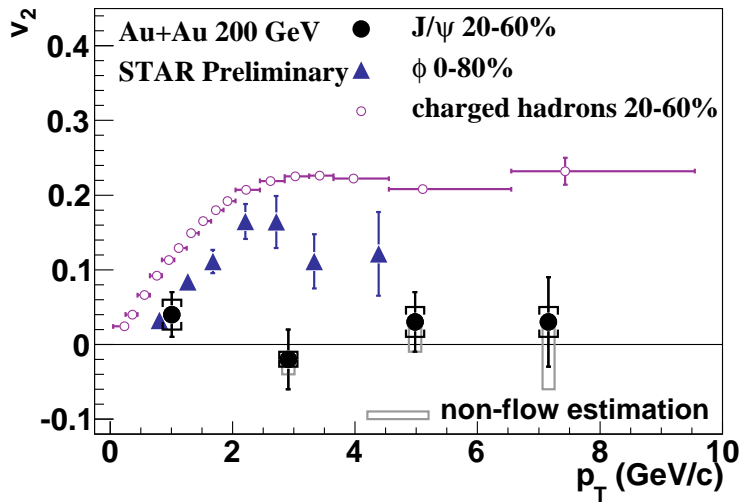}
    \end{center}
      \caption{ (Color online)
        $J/\psi$ $v_2$ (solid circles) as a functioin of $p_T$ in mid-central 20-60\% Au~+~Au 
        collisions at $\sqrt{s_{NN}}$~=~200~GeV, where open boxes show systematic uncertainties
        from non-flow effects, and brackets show those from other sources.
        Open circles show result of inclusive charged hadrons in 20-60\%~\cite{Adams:2004wz}
        and solid triangles are the $\phi$ meson $v_2$ in 0-80\%~\cite{Abelev:2007rw}
        for comparison.
      }
      \label{fig:fig3}
  \end{figure}

  It has been reported at RHIC that azimuthal anisotropy $v_2$ have shown universal scaling among
  light flavor mesons and baryons in intermediate $p_T$ range suggested by parton coalescence and 
  recombination models~\cite{Abelev:2007rw,Molnar:2003ff,Greco:2003mm,Fries:2003kq}.
  The results are strong indication of partonic degrees of freedom in the early stage of heavy ion collisions.
  Measurements of charm quark collectivity could play a potential role to answer the thermalization
  of light flavors as well as to shed more lights on the production mechanisms of charm quarks in the medium. 
  Figure~\ref{fig:fig3} shows new measurements of $J/\psi$ $v_2$ as a function of $p_T$ in 20-60\% 
  Au~+~Au collisions at $\sqrt{s_{NN}}$~=~200~GeV, up to $p_T$~=~8~GeV/$c$~\cite{Tang:QM2011}.
  Systematic errors from non-flow effects are evaluated with measurements of preliminary $J/\psi$-hadron
  correlation in p~+~p collisions by assuming that the non-flow effects are the same in p~+~p and Au~+~Au collisions.
  The $J/\psi$ $v_2$ appears to be consistent with 0 within systematic uncertainties
  for $p_T >$~2~GeV/$c$, which seems to disfavor the coalescence scenario to
  form $J/\psi$ with thermalized charm quarks.

\section{Electromagnetic probes}
\label{sec:em}

  Di-lepton productions are of particular interest among the observables 
  to explore the early stage of heavy ion collisions. They directly probe 
  the evolution of the medium and unaffected by final-state interactions
  in contrast to hadrons.

  \begin{figure}[htbp]
  \begin{tabular}{cc}
    \begin{minipage}[htbp]{0.50\textwidth}
      \includegraphics[width=\linewidth]{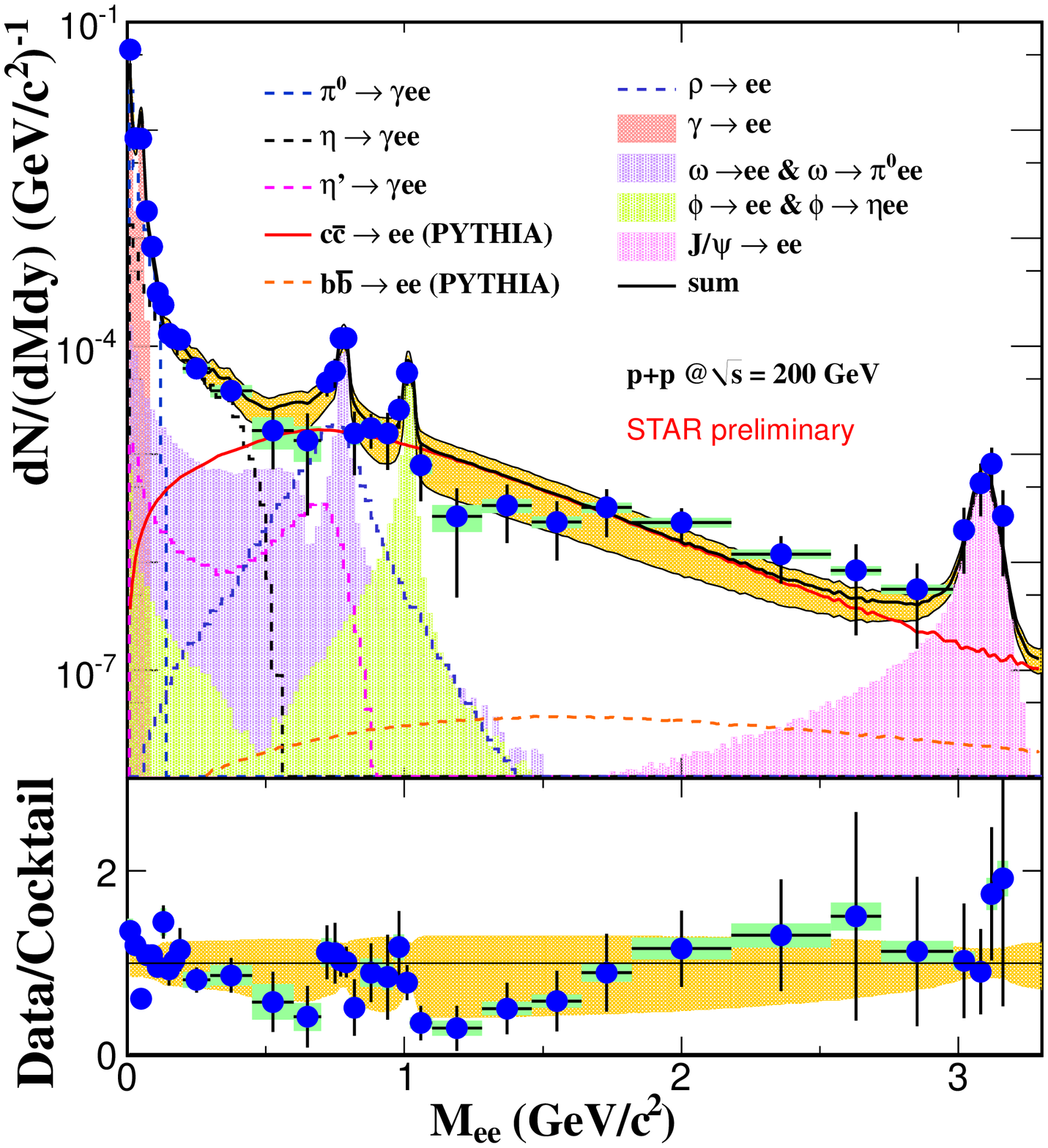}
    \end{minipage}
    \hfill
    \begin{minipage}[htbp]{0.50\textwidth}
      \includegraphics[width=\linewidth]{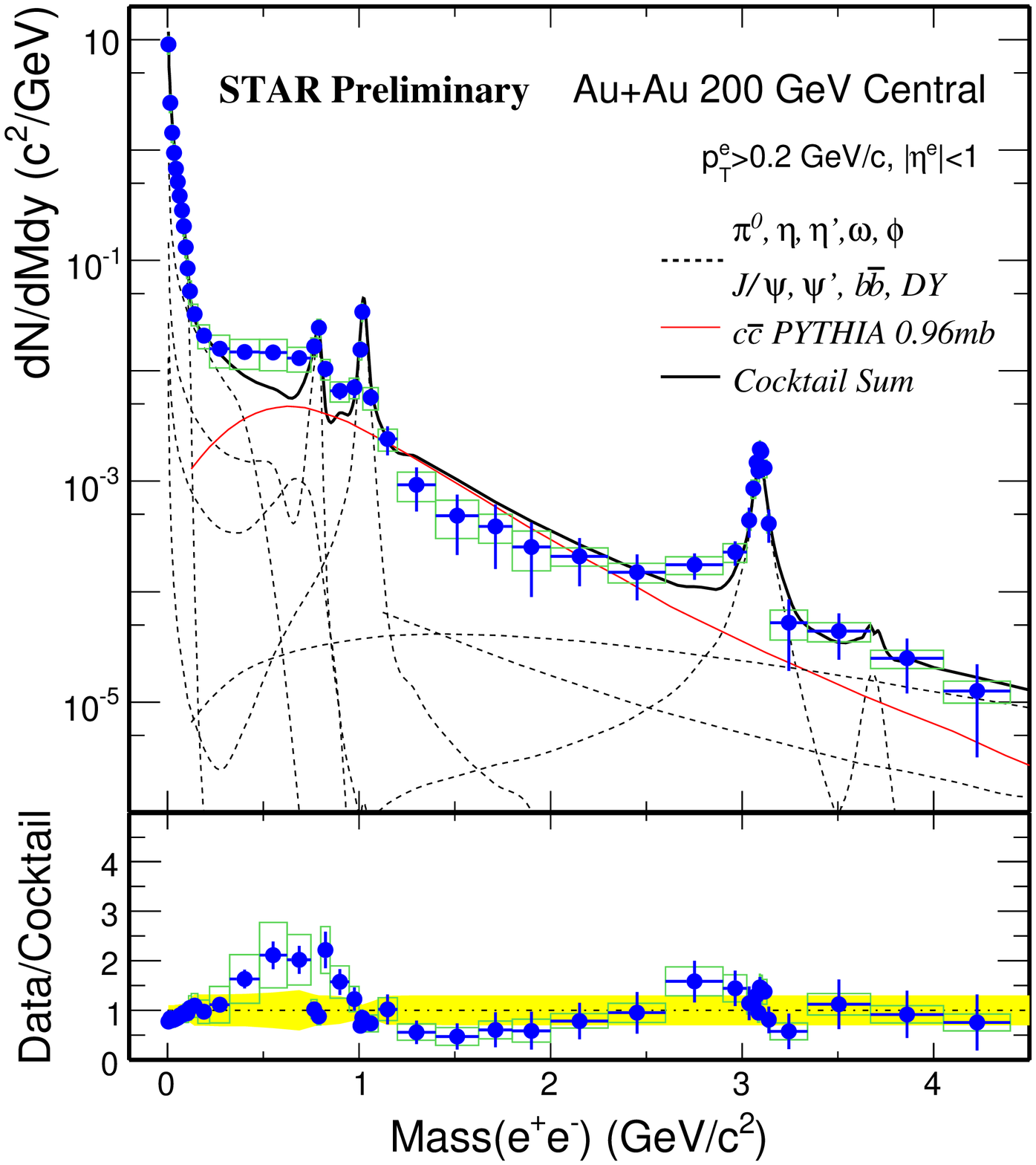}
    \end{minipage}
  \end{tabular}
      \caption{ (Color online)
        Di-electron mass spectrum in p~+~p collisions at $\sqrt{s}$~=~200~GeV (left)
        and in Au~+~Au collisions at $\sqrt{s_{NN}}$~=~200~GeV (right)
        after combinatorial background subtraction.
        Solid circles show the data, the bars denote the statistical uncertainties 
        and boxes around data points show systematic uncertainties.
        The dashed lines represent the individual 
        hadronic source used in the cocktail simulations. Bottom panels show 
        the ratio of data to cocktail simulation, where shaded bands around unity
        and the boxes around data points show the systematic uncertainties 
        on cocktail and data, respectively.
      }
      \label{fig:fig4}
  \end{figure}

  Figure~\ref{fig:fig4} presents the recent measurements of STAR di-electron mass 
  spectrum in both p~+~p (left) and Au~+~Au (right) collisions at $\sqrt{s_{NN}}$~=~200~GeV
  after combinatorial background subtraction~\cite{Zhao:QM2011}.
  The result in p~+~p collicions is consistent with the know hadronic sources 
  by cocktail simulation within systematic uncertainties,
  which provides the baseline for the measurements in Au~+~Au collisions.
  The result in central 0-5\% Au~+~Au collisions is also compared to the 
  hadronic cocktail simulation as shown in the right panel.
  The data in $M(e^+e^-) < 1$~GeV/$c^2$ appears to be enhanced by a factor of
  1.72$\pm$0.10(stat)$\pm$0.50(sys) compared to the cocktail simulation
  without $\rho^0$ contribution in the cocktail. Further study will be done 
  to study the possible modification of $\rho^0$ in the low mass region 
  and charm contributions in the intermediate mass region.

\section{Beam Energy Scan}
\label{sec:bes}

  The goal of RHIC Beam Energy Scan is to search for the QCD critical point,
  onset of signature of QGP, and softening of equation of state.
  It has been started since year 2010, and STAR experiment collected the data
  at $\sqrt{s_{NN}}$~=~7.7, 11.5 and 39 GeV in 2010, and 19.6 GeV in 2011.

  \begin{figure}[htbp]
  \begin{tabular}{cc}
    \begin{minipage}[htbp]{0.50\textwidth}
      \includegraphics[width=\linewidth]{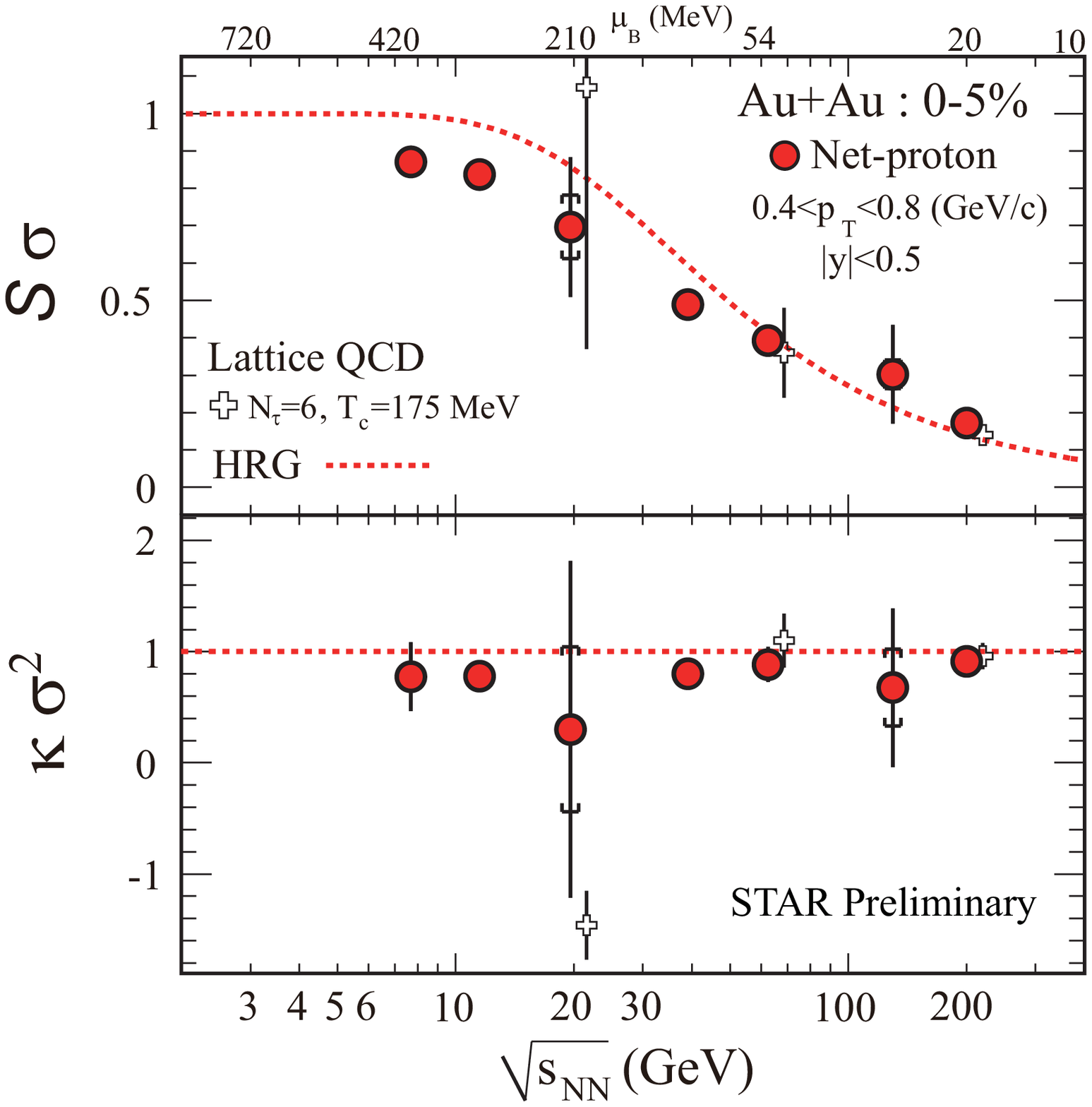}
    \end{minipage}
    \hfill
    \begin{minipage}[htbp]{0.50\textwidth}
      \includegraphics[width=\linewidth]{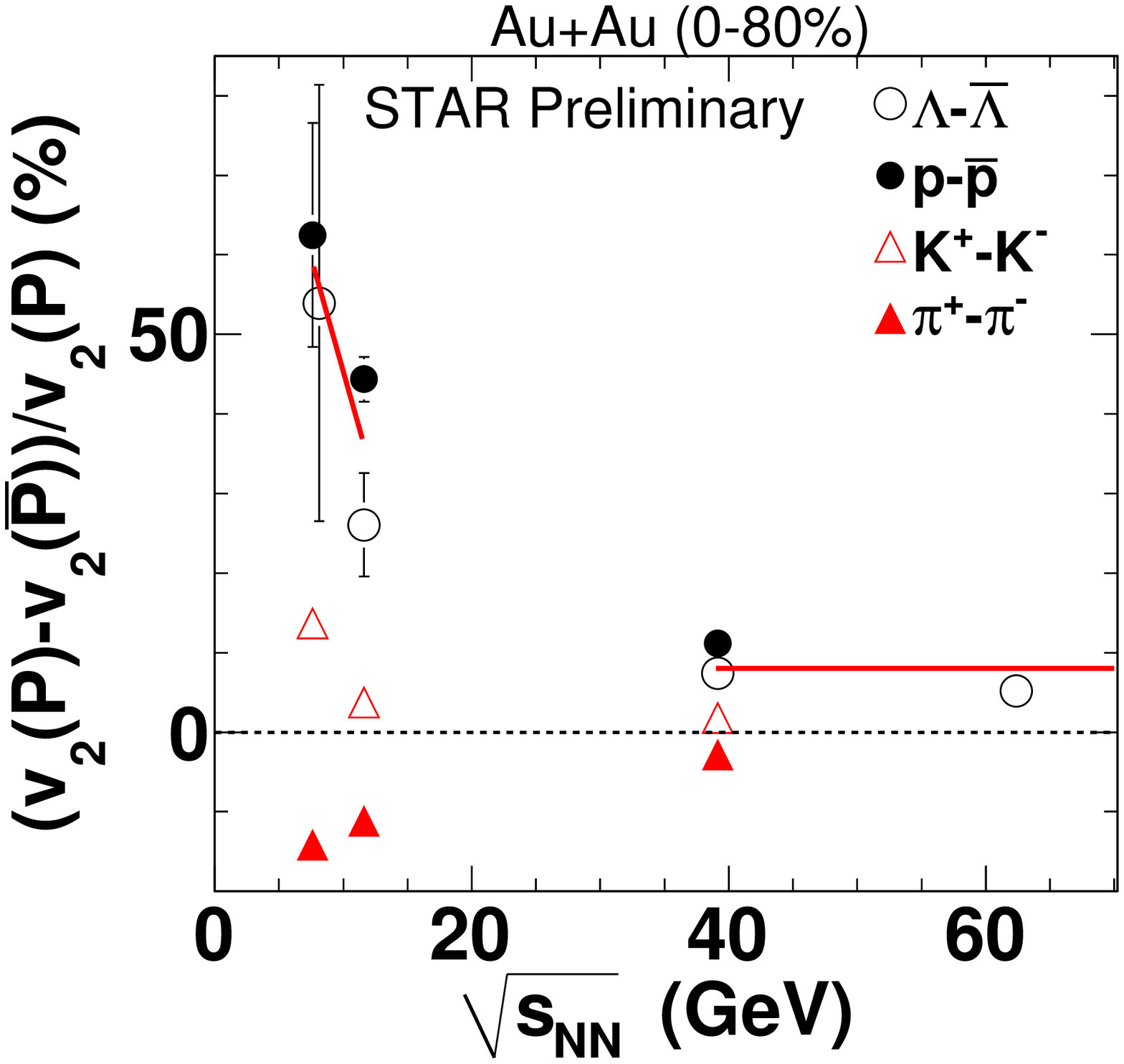}
    \end{minipage}
  \end{tabular}
      \caption{ (Color online)
        (Left)
        Product of higher moments $S\sigma$ (top) and $\kappa \sigma^2$ (bottom)
        for net-protons as a function of $\sqrt{s_{NN}}$ in the most central Au~+~Au collisions.
        Dashed lines show the calculations from Hadron Resonance Gas model~\cite{Karsch:2010ck}
        and open crosses show the Lattice QCD calculations~\cite{Gupta:2011wh}.
        (Right)
        Difference of $v_2$ between particles and anti-particles normalized
        by particle $v_2$ as a function of $\sqrt{s_{NN}}$ in 0-80\% Au~+~Au 
        collisions for $\Lambda$ (open circles), $p$ (solid circles),
        $K$ (open triangles) and $\pi$ (solid triangles). The data points 
        for $\Lambda$ at $\sqrt{s_{NN}}$~=~62.4 GeV are taken from~\cite{Abelev:2007qg}.
        Solid lines just guide eyes.
      }
      \label{fig:fig5}
  \end{figure}

  Higher moments of fluctuations for conserved quantities, such as net-baryons, net-charge and so on,
  are expected to be more sensitive to the correlation length of the system,
  and thus to be a powerful tool to search for the critical point~\cite{Stephanov:2008qz,Aggarwal:2010wy}.
  Figure~\ref{fig:fig5} (left) shows the excitation function of the product of 
  net-proton higher moments $S\sigma$ and $\kappa\sigma^2$ in the most central 
  Au~+~Au collisions, where $\sigma^2$, $S$ and $\kappa$ denote the variance,
  skewness and kurtosis, respectively~\cite{Tarnowsky:QM2011}.
  The calculations from Hadron Resonance Gas (HRG)~\cite{Karsch:2010ck}
  model as well as Lattice QCD~\cite{Gupta:2011wh} are also shown for comparison.
  The data at $\sqrt{s_{NN}}$~=~62.4-200 GeV are consistent with both HRG and 
  Lattice QCD results, while the results at $\sqrt{s_{NN}}$~=~7.7-39 GeV show
  deviation from the HRG model. The data at $\sqrt{s_{NN}}$~=~19.6~GeV show 
  the largest deviation with huge statistical uncertainty, which will be improved 
  by using the new 19.6~GeV data with much higher statistics.

  As discussed earlier, the number of quark scaling of $v_2$ could be the
  strong indication of partonic degrees of freedom at $\sqrt{s_{NN}}$~=~200~GeV. The observed 
  scaling can be used as a turn-off signature if the hadronic stage becomes dominant in 
  lower energies at BES.
  Figure~\ref{fig:fig5} (right) shows the excitation function of
  the difference of $v_2$ between particles and anti-particles 
  normalized by particle $v_2$ in 0-80\% Au~+~Au collisions at $\sqrt{s_{NN}}$~=~7.7-62.4 GeV~\cite{Schmah:QM2011}.
  The results at 62.4~GeV has shown about 5\% difference of $v_2$ between $\Lambda$ and $\bar{\Lambda}$~\cite{Abelev:2007qg}.
  At $\sqrt{s_{NN}}$~=~7.7 and 11.5 GeV, the difference of $v_2$ is significantly larger $\sim$ 50\% for baryons
  than those at higher energies. The observed difference is similar for $p$ and $\Lambda$.
  There are 5-10\% difference on meson $v_2$ as well at $\sqrt{s_{NN}}$~=~7.7 and 11.5 GeV.

\section{Summary}

  We report recent results from STAR experiment at RHIC.
  The away side jets from jet-hadron correlation in Au~+~Au collisions show significant modifications 
  as compared to that in p~+~p collisions. The results seem to be consistent with the radiative energy loss picture.
  The near side peak obtained by di-hadron correlation with identified $\pi$ trigger is
  higher than that for $p+K$ trigger, which is similar in both d~+~Au and Au~+~Au collisioins.
  This result appears to disfavor the expectations from naive coalescence of thermal partons.
  The high $p_T$ $J/\psi$ shows stronger suppression in more central collisions, while the 
  magnitude of suppression is less compared to the results in lower $p_T$. The suppression of $\Upsilon$
  is observed at 3~$\sigma$ level including p~+~p uncertainties at most central 0-10\% collisions.
  The $J/\psi$ $v_2$ appears to be consistent with 0 in $2 < p_T < 8$~GeV/$c$, which disfavors 
  the scenario of naive coalescence for thermal charm quarks.
  The STAR new measurement of di-electron mass spectrum in Au~+~Au collisions 
  at $\sqrt{s_{NN}}$~=~200~GeV is presented compared to that in p~+~p collisions.
  The result in p~+~p collisions is consistent with hadronic cocktail simulation, which provides 
  the baseline for the measurements in Au~+~Au collisions. The enhancement of di-electron 
  spectrum is observed below $M(e^+e^-) <$~1~GeV/$c^2$ in Au~+~Au collisions. Further investigation 
  will be carried out to understand the possible modification of vector mesons as well as charm 
  contributions.
  Excitation function of higher moments of net-protons are reported. The results are in good agreement 
  with HRG model at higher energies, while they seems to deviate from the HRG model in $\sqrt{s_{NN}}$~=~7.7-39~GeV.
  Significant difference between baryon and anti-baryon $v_2$ is observed for $p$ and $\Lambda$
  in $\sqrt{s_{NN}}$~=~7.7 and 11.5~GeV. Further studies will be performed with 19.6 and 27~GeV data
  in order to search for the QCD critical point and the onset of QGP.


\section*{References}
  \begin{thebibliography}{99}

  \bibitem{Adams:2003kv}
    J.~Adams {\it et al.}  [STAR Collaboration],
      Phys.\ Rev.\ Lett.\  {\bf 91}, 172302 (2003)

  \bibitem{Adler:2002tq}
    C.~Adler {\it et al.}  [STAR Collaboration],
      Phys.\ Rev.\ Lett.\  {\bf 90}, 082302 (2003)

  \bibitem{Alice:QM2011}
   A.~Ohlson [STAR Collaboration],
   these proceedings

  \bibitem{Kolja:QM2011}
   K.~Kauder [STAR Collaboration],
   these proceedings

  \bibitem{Matsui:1986dk}
   T.~Matsui and H.~Satz,
   Phys.\ Lett.\  B {\bf 178}, 416 (1986).



   \bibitem{Adare:2006ns}
     A.~Adare {\it et al.}  [PHENIX Collaboration],
    Phys.\ Rev.\ Lett.\  {\bf 98}, 232301 (2007)

   \bibitem{Abelev:2009qaa}
     B.~I.~Abelev {\it et al.}  [STAR Collaboration],
     Phys.\ Rev.\  C {\bf 80}, 041902 (2009)

   \bibitem{Grandchamp:2005yw}
     L.~Grandchamp, S.~Lumpkins, D.~Sun, H.~van Hees and R.~Rapp,
     Phys.\ Rev.\  C {\bf 73}, 064906 (2006)

   \bibitem{Lin:2000ke}
     Z.~w.~Lin and C.~M.~Ko,
     Phys.\ Lett.\  B {\bf 503}, 104 (2001)
             %

    \bibitem{Tang:2011xq}
      Z.~Tang {\it et al.},
    arXiv:1101.1912 [nucl-ex].

    \bibitem{Tang:QM2011}
      Z.~Tang [STAR Collaboration],
      these proceedings

    \bibitem{Reed:QM2011}
      R.~Reed [STAR Collaboration],
      these proceedings

    \bibitem{Adams:2004wz}
      J.~Adams {\it et al.}  [STAR Collaboration],
      Phys.\ Rev.\ Lett.\  {\bf 93}, 252301 (2004)

    \bibitem{Abelev:2007rw}
      B.~I.~Abelev {\it et al.}  [STAR Collaboration],
      Phys.\ Rev.\ Lett.\  {\bf 99}, 112301 (2007)

    \bibitem{Molnar:2003ff}
      D.~Molnar and S.~A.~Voloshin,
      Phys.\ Rev.\ Lett.\  {\bf 91}, 092301 (2003)

    \bibitem{Greco:2003mm}
      V.~Greco, C.~M.~Ko and P.~Levai,
      Phys.\ Rev.\  C {\bf 68}, 034904 (2003)

    \bibitem{Fries:2003kq}
      R.~J.~Fries, B.~Muller, C.~Nonaka and S.~A.~Bass,
      Phys.\ Rev.\  C {\bf 68}, 044902 (2003)

    \bibitem{Zhao:QM2011}
      J.~Zhao [STAR Collaboration],
      these proceedings


     \bibitem{Stephanov:2008qz}
       M.~A.~Stephanov,
       Phys.\ Rev.\ Lett.\  {\bf 102}, 032301 (2009)

      \bibitem{Aggarwal:2010wy}
        M.~M.~Aggarwal {\it et al.}  [STAR Collaboration],
        Phys.\ Rev.\ Lett.\  {\bf 105}, 022302 (2010)

    \bibitem{Tarnowsky:QM2011}
      T.~Tarnowsky [STAR Collaboration],
      these proceedings


    \bibitem{Karsch:2010ck}
      F.~Karsch and K.~Redlich,
      Phys.\ Lett.\  B {\bf 695}, 136 (2011)


    \bibitem{Gupta:2011wh}
      S.~Gupta, X.~Luo, B.~Mohanty, H.~G.~Ritter and N.~Xu,
       Science {\bf 332}, 1525 (2011)

    \bibitem{Schmah:QM2011}
      A.~Schmah [STAR Collaboration],
      these proceedings

    \bibitem{Abelev:2007qg}
      B.~I.~Abelev {\it et al.}  [the STAR Collaboration],
      Phys.\ Rev.\  C {\bf 75}, 054906 (2007)
  \end{thebibliography}
\end{document}